\documentstyle[epsfig,twoside,fleqn,espcrc2,axodraw]{article}

\def\lsim{\raise0.3ex\hbox{$\;<$\kern-0.75em\raise-1.1ex\hbox{$\sim\;$}}}
\def\gsim{\raise0.3ex\hbox{$\;>$\kern-0.75em\raise-1.1ex\hbox{$\sim\;$}}}

\def\etal{{\it et al.}}
\def\half{{\textstyle{1 \over 2}}}

\def\fourth{{\textstyle{1 \over 4}}}

\def\bold#1{\setbox0=\hbox{$#1$}
     \kern-.025em\copy0\kern-\wd0
     \kern.05em\copy0\kern-\wd0
     \kern-.025em\raise.0433em\box0 }

\newcommand{\AmS}{{\protect\the\textfont2
  A\kern-.1667em\lower.5ex\hbox{M}\kern-.125emS}}

\begin{document}


\title{
\vskip-1.3cm\rightline{\small{hep-ph/9909462}}
\vskip-0.2cm\rightline{\small{UCCHEP/3-99}}
\vskip-0.2cm\rightline{\small{FSU-HEP-990723}}
\vskip+0.1cm
Gauge and Yukawa Unification in SUSY with Bilinearly Broken 
R--Parity}

\author{    Marco Aurelio D\'\i az
   \address{
            Facultad de F\'\i sica,
            Universidad Cat\'olica de Chile, 
            Av. Vicu\~na Mackenna 4860, Santiago, Chile, and \\
\hskip0.3cm Department of Physics,
            Florida State University, 
            Tallahassee, Florida 32306, USA.
\\ \vskip0.4cm Talk given at the International Workshop on Particles in 
            Astrophysics and Cosmology: From Theory to Observation,
            Valencia, Spain, May 3-8, 1999.
}}

\begin{abstract}
In a supersymmetric model where R--Parity and lepton number are violated 
bilinearly in the superpotential, which can explain the solar and 
atmospheric neutrino problems, we study the unification of gauge and Yukawa
couplings at the GUT scale. We show that bottom--tau Yukawa coupling
unification can be achieved at any value of $\tan\beta$, and that the
strong coupling constant prediction from unification of gauge couplings 
is closer to the experimental value compared with the MSSM. We also study the 
predictions for $V_{cb}$ in a Yukawa texture ans\"atze.
\end{abstract}

\maketitle

Bilinear R--Parity Violation (BRpV) has attracted a lot of attention lately
\cite{brpvnew,brpvus} due to its prediction for neutrino masses 
\cite{SponRpB,rpre} in connection with the recent results from 
SuperKamiokande \cite{SKK}, confirming the deficit of muon neutrinos from
atmospheric neutrino data \cite{expothers}. The simplest interpretation of 
the data is in terms of $\nu_{\mu}$ to $\nu_{\tau}$ flavour oscillations with
maximal mixing and a mass squared difference of
\begin{equation}
\Delta m^2_{\mathrm{atm}}\approx 10^{-3}-10^{-2}\,\,{\mathrm{eV}}^2\,.
\label{mdiffatm}
\end{equation}
In addition, the solar neutrino experiments \cite{solar} imply the 
existence of another independent neutrino mass squared difference
\begin{eqnarray}
\Delta m^2_{\mathrm{sun}}&\approx& 10^{-10}\,\,{\mathrm{eV}}^2\,(\mathrm{VO})
\nonumber\\
\Delta m^2_{\mathrm{sun}}&\approx& 10^{-6}-10^{-4}\,\,{\mathrm{eV}}^2\,
(\mathrm{MSW})
\label{mdiffsun}
\end{eqnarray}
where VO stands for Vacuum Oscillation solution and MSW for the 
Mikheyev--Smirnov--Wolfenstein solution \cite{MSW}. The solar effect could
be due to $\nu_e$ to $\nu_{\mu}$ oscillations.

The solar and atmospheric neutrino problems can be solved in the context of
BRpV \cite{RDHPV} where the superpotential 
\begin{equation}
W=W_{\mathrm{Yuk}}-\mu\widehat H_d\widehat H_u+\epsilon_i\widehat L_i
\widehat H_u
\label{superpot}
\end{equation}
contains three terms that mix the three lepton superfields with the Higgs
superfield responsible for the up--type quark masses. The three mixing terms
violate R--Parity and lepton number and are proportional to parameters
$\epsilon_i$ with units of mass.

The three neutrinos mix with the four neutralinos, and in a see-saw type of 
mechanism one of them acquire a mass at tree level and the other two remain 
degenerate and massless. In the case of $\epsilon_1\ll\mu,M_{1/2}$ the tree
level neutrino mass can be approximated to \cite{RDHPV,HirschV}
\begin{equation}
m_{\nu}\approx{{g^2M_1+g'^2M_2}\over{4{\mathrm{det}}({\bf M}_{\chi^0})}}
|\vec\Lambda|^2\,,
\label{numass}
\end{equation}
where $\Lambda_i=\mu v_i+v_d\epsilon_i$ and $v_i$ are the vev's of the
sneutrinos. The matrix ${\bf M}_{\chi^0}$ is the $4\times4$ submatrix
corresponding to the original neutralinos. It can be shown that the 
parameters $\Lambda_i\approx\mu'v'_i$ are directly proportional to the 
sneutrino vev's $v'_i$ in the basis where the $\epsilon_i$ terms have been 
removed from the superpotential. 

In order to calculate reliable neutrino masses and mixings it is imperative
to include one--loop corrections to the three generations
\cite{RDHPV,Hempfling}. In this way, the degeneracy and masslessness of 
the lightest two neutrinos is lifted. The renormalized mass matrix has the 
form
\begin{eqnarray}
M_{ij}^{pole} & =& M_{ij}^{\overline {DR}}(\mu_R) 
              + \frac{1}{2} \Big(
                \Pi_{ij}(p_i^2) + \Pi_{ij}(p_j^2) \nonumber\\
              &  - & m_{\chi^0_i} \Sigma_{ij}(p_i^2) - 
                  m_{\chi^0_j} \Sigma_{ij}(p_j^2) \Big)\,.
\label{1loopMass}
\end{eqnarray}
where $\mu_R$ is an arbitrary scale and $\Pi_{ij}$ and $\Sigma_{ij}$ are
self energies. The explicit scale dependence of the self energies is canceled
by the implicit scale dependence of the tree level masses in the 
$\overline{DR}$ scheme. The averaged form of the renormalized mass matrix is
necessary for explicit gauge independence. The loops include:

\begin{picture}(180,40)(0,20) 
\ArrowLine(0,25)(25,25)
\ArrowArcn(40,25)(15,180,0)
\DashCArc(40,25)(15,180,0){3}
\ArrowLine(55,25)(80,25)
\Text(90,25)[]{$+$}
\ArrowLine(100,25)(125,25)
\ArrowArcn(140,25)(15,180,0)
\PhotonArc(140,25)(15,180,0){2}{8.5}
\ArrowLine(155,25)(180,25)
\Text(12,32)[]{$\nu$}
\Text(67,32)[]{$\nu$}
\Text(112,32)[]{$\nu$}
\Text(167,32)[]{$\nu$}
\Text(43,48)[]{$q,F^+,F^0$}
\Text(43,3)[]{$\tilde q,S^-,S^0$}
\Text(142,48)[]{$F^+, F^0$}
\Text(141,3)[]{$W^-, Z$}
\end{picture}
\vskip 0.8cm
\noindent
where $F^+$ are mixtures of charginos and charged leptons, $F^0$ are 
neutralinos and neutrinos, $S^+$ are charged Higgs and charged sleptons, 
and $F^0$ are neutral Higgs and sneutrinos. 

We work in the general $R_{\xi}$ gauge, and to achieve explicit gauge
invariance we need to include the tadpole graph for the Goldstone 
bosons into the self energies. There are five tadpole equations associated
to the real parts of the two neutral Higgs and three sneutrinos:
\begin{equation}
V_{linear}=t_d Re(H^0_d)+t_u Re(H^0_u)+t_i Re(\tilde\nu_{i})\,.
\label{Vlinear}
\end{equation}
The renormalized tadpoles are
\begin{equation}
t_{\alpha}=t^0_{\alpha} -\delta t^{\overline{DR}}_{\alpha}
+T_{\alpha}(Q)=t^0_{\alpha} +T^{\overline{DR}} _{\alpha}(Q)
\label{tadpoles}
\end{equation}
where $T^{\overline{DR}} _{\alpha}(Q)\equiv -\delta
t^{\overline{DR}}_{\alpha} +T_{\alpha}(Q)$ are the finite one--loop
tadpoles without the Goldstone contribution. 

\begin{figure}[htb]
\vspace{-5.7cm}
\centerline{ \hskip 2.5cm \epsfxsize 4.0 truein \epsfbox {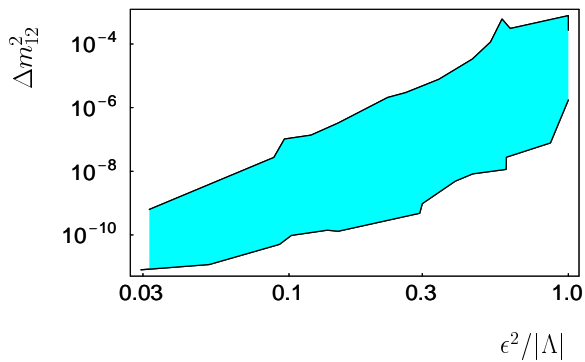}} 
\vskip 0.4cm
\caption{Solar mass squared difference $\Delta m^2_{12}$ as a function of 
$\epsilon^2/|\Lambda|$. Lower values of $\Delta m^2_{12}$ corresponds to
VO solution and high values to the MSW effect.}
\vskip -0.6cm
\label{lettfig3}
\end{figure}
The solar mass squared difference $\Delta m^2_{\mathrm{sun}}=\Delta m^2_{12}$
is plotted in Fig.~\ref{lettfig3} as a function of $\epsilon/|\Lambda|$,
where $\epsilon^2=\sum_i\epsilon_i^2$ and $\Lambda^2=\sum_i\Lambda_i^2$.
Lower values of $\epsilon/|\Lambda|$ leads to VO solutions and large
values to solutions with the MSW effect. Maximality of the atmospheric 
angle is found for $|\Lambda_{\mu}|\approx|\Lambda_{\tau}|$, and maximality 
of the solar angle is obtained if $\epsilon_e\approx\epsilon_{\mu}$, as
long as $\Lambda_e$ is about a decade smaller than the other two. 

An important consequence of the supersymmetric solutions to the neutrino 
problems is the necessity of $\tan\beta\lsim10$, implying that the lightest
Higgs boson mass satisfy $m_h\lsim115$ GeV. In BRpV the neutral Higgs mix
with the sneutrinos, and although this mixing does not affect the upper
bound on $m_h$, it can reduce the mass in a few GeV \cite{DRV}. A large
part of the Higgs mass $m_h$ comes from radiative corrections near 
$\tan\beta=1$ \cite{DHaberii} and therefore it is smaller than at high 
$\tan\beta$. This is the reason why LEP2 has started to prove this region
of parameter space, preliminary ruling out $1\le\tan\beta\lsim 1.8$
\cite{LEPmh}.

The analysis of BRpV with three massive neutrinos is very involved and
for many applications it is enough to consider the one--generation 
approximation. In the study of gauge and Yukawa unification, the details
of neutrino masses and mixing are not relevant, and for simplicity we 
consider BRpV only in the tau sector. The superpotential is the one in 
eq.~(\ref{superpot}) with $\epsilon_1=\epsilon_2=0$ and $\epsilon_3\ne0$.
in addition, an extra soft parameter $B_3$ is introduced
\begin{equation}
V_{\mathrm{soft}}^{\mathrm{BRpV}}=V_{\mathrm{soft}}^{\mathrm{MSSM}}+
B_3\epsilon_3\tilde L_3H_2+h.c.
\label{softnew}
\end{equation}
In this context, the tau neutrino acquire a mass at tree level given by
\begin{equation}
m_{\nu_{\tau}}\approx{{g^2}\over{2M}}v'^2_3
\label{taunumass}
\end{equation}
where $v'_3$ is the sneutrino vev in the basis where the $\epsilon_3$ term is 
removed from the superpotential. The tadpole equations allow us to find an
approximated expression for this vev
\begin{equation}
v'_3\approx -{{\epsilon_3\mu}\over{\mu'^2m_{\tilde\nu^0_{\tau}}^2}}
\left(v'_1\Delta m^2+\mu'v_2\Delta B\right)
\label{sneuvev}
\end{equation}
where $\Delta m^2=m_{H_1}^2-m_{L_3}^2$ and $\Delta B=B_3-B$ are evaluated at 
the weak scale. In models with universal boundary conditions at the GUT scale,
these two differences at the weak scale are radiativelly generated and 
proportional to the bottom Yukawa coupling squared.

The sneutrino vev $v_3$ contribute to the $W$ boson mass 
$m_W^2=\fourth g^2(v_u^2+v_d^2+v_3^2)$, therefore the Higgs vevs are smaller
compared to the MSSM case. Because of this, although the relation between 
the quark masses and the Yukawa couplings does not change
\begin{equation}
m_t^2=\half h_t^2v_u^2\,,\qquad m_b^2=\half h_b^2v_d^2\,,
\label{quarkmasses}
\end{equation}
the numerical value of the Yukawas is different compared with the MSSM. 
The case of the tau Yukawa coupling is different due to the tau mixing
with the charginos. The tau mass is
\begin{equation}
m_{\tau}^2=\half h_{\tau}^2v_d^2(1+\delta)
\label{taumass}
\end{equation}
where $\delta\ge0$ depends on the parameters of the chargino--tau mass matrix
\cite{textura}.

\begin{figure}[htb]
\vspace{-0.7cm}
\centerline{ \hskip 0.cm \epsfxsize 3.0 truein \epsfbox {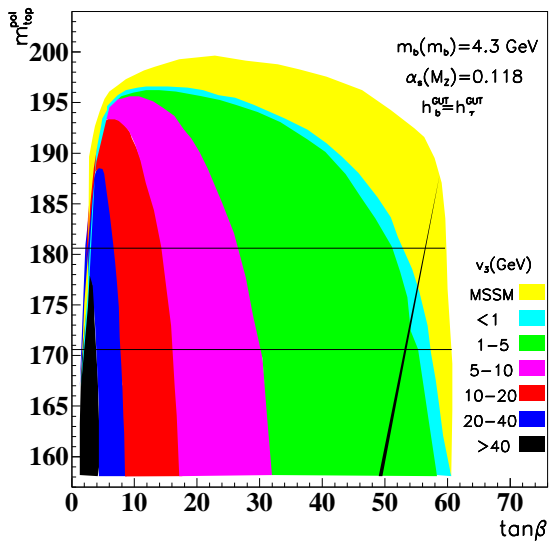}} 
\vskip -0.8cm
\caption{Top quark pole mass as a function of $\tan\beta$ for different
values of the sneutrino vev $v_3$. Bottom--tau Yukawa unification in 
BRpV can be obtained at any value of $\tan\beta$. Top--bottom--tau 
unification (inclined line) is achieved at high $\tan\beta$ in a slightly 
wider region compared to the MSSM.}
\vskip -0.4cm
\label{aretop}
\end{figure}
In this context we have made a complete scan over parameter space looking 
for solutions with bottom--tau Yukawa unification within $1\%$. In 
Fig.~\ref{aretop} we plot the top quark pole mass as a function of 
$\tan\beta$ differenciating the regions with the value of the sneutrino 
vev $v_3$ (there is some overlap between the regions that we do not show). 
The two horizontal lines correspond to the $1\sigma$ experimental 
determination of the top quark mass \cite{topex}. The MSSM case is also 
shown and the usual two solutions, one for large $\tan\beta\sim 55$ and 
one for small $\tan\beta\sim 2$, can be observed. As we can infer from the 
figure, bottom--tau Yukawa unification in BRpV can be achieved at any value 
of $\tan\beta$ provided we tune the value of $v_3$. In addition, 
top--bottom--tau Yukawa unification (inclined line) is achieved in a 
slightly wider region at high $\tan\beta$ \cite{yukunif}.

To understand this result, consider the ratio between $h_b$ and $h_{\tau}$
at the weak scale. According to eqs.~(\ref{quarkmasses})
and (\ref{taumass}) this ratio is
\begin{equation}
{{h_b}\over{h_{\tau}}}(m_{weak})={{m_b}\over{m_{\tau}}}\sqrt{1+\delta}
\label{ratio1}
\end{equation}
with $\delta$ increasing when $v_3$ departs from zero. In addition, solving 
the RGE's with bottom--tau unification at the weak scale we get
\begin{eqnarray}
&&\!\!\!\!\!\!\!\!\!\!\!\!
{{h_b}\over{h_{\tau}}}(m_{weak})\approx
\nonumber\\
&&\!\!\!\!\!\!\!\!\!\!\!\!
exp\left[{1\over{16\pi^2}}\left({16\over3}g_s^2-3h_b^2-h_t^2\right)
\ln{{M_{GUT}}\over{m_{weak}}}\right]
\label{hb_htau}
\end{eqnarray}
Comparing eqs.~(\ref{ratio1}) and (\ref{hb_htau}) we infer that the 
combination $3h_b^2+h_t^2$ decreases when $v_3$ departs from zero ($v_3=0$
corresponds to the MSSM).

\begin{figure}[htb]
\vspace{-1.2cm}
\centerline{ \hskip .6cm \epsfxsize 2.4 truein \epsfbox {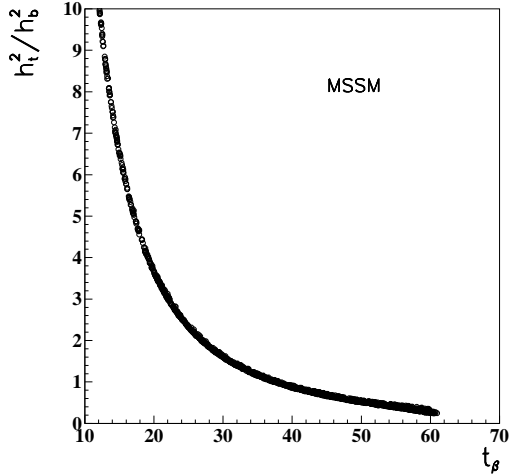}} 
\vskip -1.7cm
\caption{Ratio between the top and bottom quark Yukawa couplings at the
weak scale in the MSSM as a function of $\tan\beta$.}
\vskip -0.6cm
\label{hthbmssm}
\end{figure}
In Fig.~\ref{hthbmssm} we plot the ratio $h_t/h_b$ as a function of 
$\tan\beta$ for the MSSM points corresponding to the outer band in 
Fig.~\ref{aretop}. The points with acceptable $m_t$ (within $1\sigma$) and
high $\tan\beta$ lie in the range $53\lsim\tan\beta\lsim60$ and in this
reagion clearly $h_b$ dominates in the combination $3h_b^2+h_t^2$. If
$v_3$ departs from zero (away from the MSSM) then $h_b$ decreases in order
to achieved bottom-tau unification. In order to keep the bottom quark
mass constant, the vev $v_d$ increases, and to keep the gauge boson masses
constant the vev $v_2$ decreases, implying that unification is achieved
in BRpV at smaller values of $\tan\beta$.
\begin{figure}[htb]
\vspace{-1.2cm}
\centerline{ \hskip .6cm \epsfxsize 2.4 truein \epsfbox {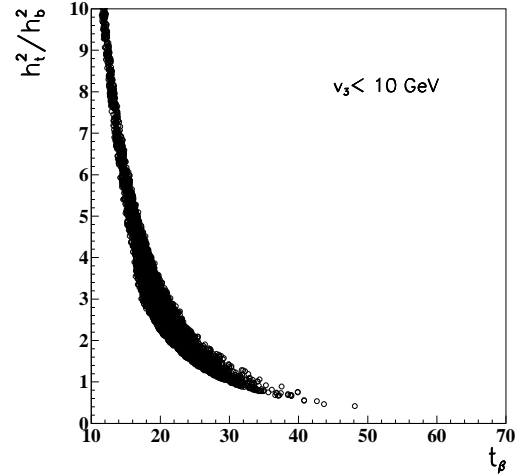}} 
\vskip -1.7cm
\caption{Ratio between the top and bottom quark Yukawa couplings at the
weak scale in the BRpV as a function of $\tan\beta$ for $5<\tan\beta<10$
GeV.}
\vskip -0.6cm
\label{hthb10}
\end{figure}
In Fig.~\ref{hthb10} we have the ratio $h_t/h_b$ as a function of $\tan\beta$
for BRpV with $5<v_3<10$ GeV. Looking at the figure we confirm that the
region where $h_b$ dominates over $h_t$ in eq.~(\ref{hb_htau}) is at smaller
values of $\tan\beta$, where $b-\tau$ unification and correct $m_t$ is
achieved.

It is clear that bottom--tau unification in BRpV is controlled by the sneutrino
vev $v_3$ and not directly by the neutrino masses. It is perfectly possible
to have large effects in bottom--tau unification and a small tau neutrino 
mass, although the complete case with three neutrinos and masses calculated 
up to one--loop is under investigation.

\begin{figure}[htb]
\vspace{-0.7cm}
\centerline{ \hskip .1cm \epsfxsize 3.2 truein \epsfbox {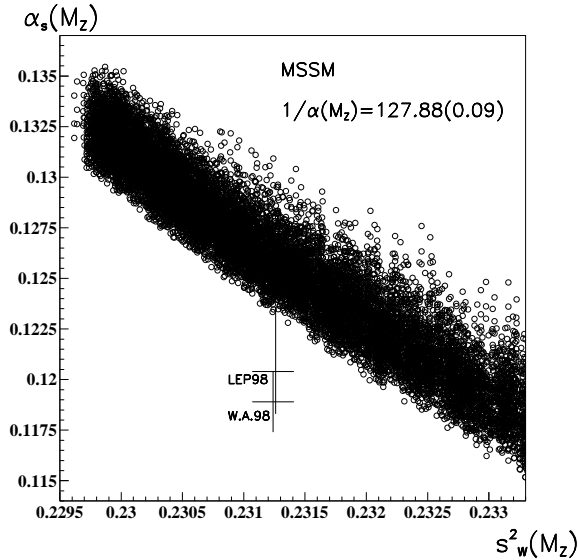}} 
\vskip -0.5cm
\caption{Gauge coupling GUT unification prediction for $\alpha_s(m_Z)$
as a function of the weak mixing angle $\sin^2\theta_W(m_Z)$ in the MSSM.}
\vskip -0.5cm
\label{mssm}
\end{figure}

Another interesting effect controlled, as we will see below, by the sneutrino
vev $v_3$ are the predictions for $\alpha_s$ from unification of gauge
couplings at the GUT scale \cite{DFRV}. The experimental world average of the
strong coupling constant $\alpha_s(m_Z)^{W.A.}=0.1189\pm0.0015$ \cite{wa}
is about $2\sigma$ lower than the GUT prediction in the MSSM \cite{alfamssm}, 
as we illustrate in Fig.~\ref{mssm}.
In this figure we have made a scan over all parameter space in the 
MSSM--SUGRA including supersymmetric threshold corrections given by 
\cite{alfamssm}
\begin{equation}
\Delta \alpha_s^{SUSY}= -\frac{19\alpha_s^2}{28\pi}
\ln\left(\frac{T_{SUSY}}{\hbox{M}_t}\right)
\label{threshSUSY}
\end{equation}
where $T_{SUSY}$ is an effective mass scale given by
\begin{eqnarray}
&&\!\!\!\!\!\!\!\!\!\!\!\!
T_{SUSY}=m_{\widetilde H}
\left({{m_{\widetilde W}}\over{m_{\tilde g}}}\right)^{28\over 19}\Bigg[
\nonumber\\
&&\!\!\!\!\!\!\!\!\!\!\!\!
\left({{m_{\tilde l}}\over{m_{\tilde q}}}\right)^{3\over 19}
\left({{m_H}\over{m_{\widetilde H}}}\right)^{3\over 19}
\left({{m_{\widetilde W}}\over{m_{\widetilde H}}}\right)^{4\over 19}
\Bigg]\,.
\label{TSUSY}
\end{eqnarray}
The $2\sigma$ difference is not a real discrepancy, nevertheless, it is 
interesting to compare it with the predictions for $\alpha_s$ in BRpV.

\begin{figure}[htb]
\vspace{-0.7cm}
\centerline{ \hskip .1cm \epsfxsize 3.2 truein \epsfbox {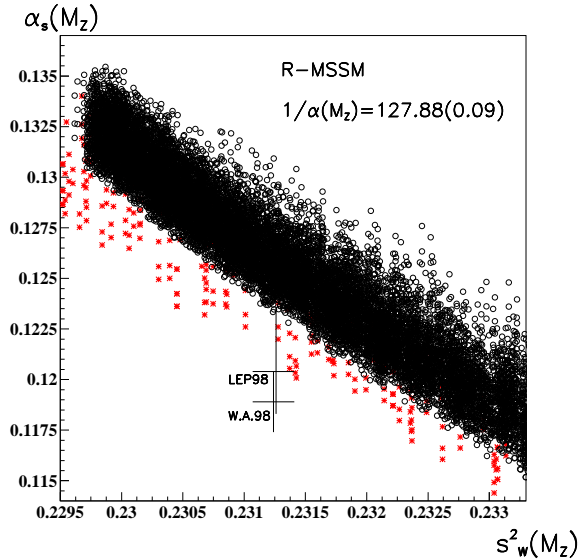}} 
\vskip -0.5cm
\caption{Gauge coupling GUT unification prediction for $\alpha_s(m_Z)$
as a function of the weak mixing angle $\sin^2\theta_W(m_Z)$ in BRpV.}
\vskip -0.7cm
\label{rmssm}
\end{figure}
With BRpV embedded into SUGRA with universality of soft masses at the GUT 
scale, where the gauge couplings unify, we have made a scan over parameter 
space looking for the prediction of $\alpha_s$ at the weak scale. The 
results are presented in Fig.~\ref{rmssm} where we plot the strong 
coupling constant as a function of $\sin^2\theta_W(m_Z)$. Interestingly, 
there is a $1\sigma$ improvement compared to the MSSM. This effect can be 
understood by noticing that the Yukawa couplings, which contribute to the 
running of $\alpha_s$ at the two--loop level, make a contribution to
$\alpha_s$ that can be approximated by
\begin{equation}
\Delta\alpha_s^{YUK} \approx -\frac{\alpha_s^2}{32\pi^3}
\ln\left(\frac{\hbox{M}_{U}}{\hbox{M}_t}\right)
b_{3}^{\prime}\left\{h^2_t + h^2_b\right\}\,,
\label{DYuktoalfas}
\end{equation}
and the difference between BRpV and MSSM is that in the former case the
combination $h_t^2+h_b^2$ can be larger than in the later case. From 
eq.~(\ref{quarkmasses}) we get
\begin{equation}
h_t^2+h_b^2=2\left({{m_t^2}\over{v_u^2}}+{{m_b^2}\over{v_d^2}}\right)\,.
\label{hthb}
\end{equation}
In the MSSM at high values of $\tan\beta$ both Yukawa couplings are 
comparable, the vev $v_u$ is very close to 246 GeV, and $v_d$ is just a few
GeV. What happens in BRpV is that sneutrino vevs $v_3$ of only a few GeV 
(comparable to $v_d$) can lift the value of the bottom Yukawa coupling to
twice as large, since to keep the gauge boson masses constant $v_d$ must
decrease.

\begin{figure}[htb]
\vspace{-1.5cm}
\centerline{ \hskip .6cm \epsfxsize 2.8 truein \epsfbox {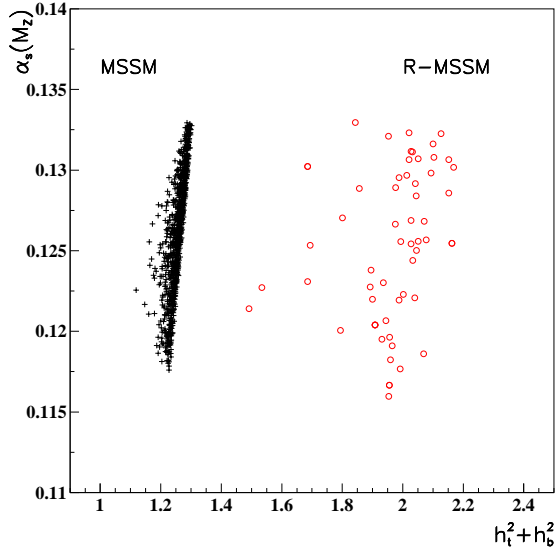}} 
\vskip -1.7cm
\caption{Strong coupling constant as a function of $h_t^2+h_b^2$ for points
with lowest $\alpha_s$ for each value of $\sin^2\theta_W$ in both the MSSM 
and BRpV.}
\vskip -0.7cm
\label{alfyuk}
\end{figure}
This effect can be seen in Fig.~\ref{alfyuk} where we plot the value of
$\alpha_s(m_Z)$ as a function of the combination $h_t^2+h_b^2$ for the
points in Figs.~\ref{mssm} and \ref{rmssm} that lie in the lower part of
each strip (the smallest values of $\alpha_s$ for each value of 
$\sin^2\theta_W$). We clearly observe the effect that $h_t^2+h_b^2$ is
larger in BRpV than in the MSSM, reason why the BRpV prediction of 
$\alpha_s$ is closer to the experimental value.

\begin{figure}[htb]
\vspace{0.cm}
\centerline{ \hskip 0.cm \epsfxsize 3.0 truein \epsfbox {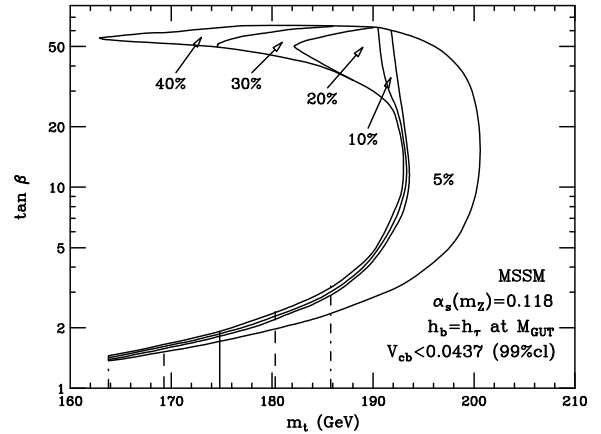}} 
\vskip -0.7cm
\caption{Regions with acceptable values of $V_{cb}$ in the 
$\tan\beta-m_{top}$ plane for the MSSM.}
\vskip -0.7cm
\label{tl99Mr}
\end{figure}
The prediction of $V_{cb}$ in the simplest Yukawa texture ans\"atze 
\cite{textura} has been also studied. In this case, in addition to the
GUT condition $h_b(M_{GUT})=h_{\tau}(M_{GUT})$ we impose the condition
\begin{equation}
\Big|V_{cb}(M_{GUT})\Big|=\sqrt{{h_c(M_{GUT})}\over{h_t(M_{GUT})}}
\label{VcbGUT}
\end{equation}
which is a prediction of the texture. We look for solutions satisfying the
experimental constraint $0.036<|V_{cb}|<0.042$ at $90\%$ c.l. \cite{expvcb} .
Solutions in the MSSM are in Fig.~\ref{tl99Mr} imposing the two boundary 
conditions at the GUT scale within the percentage indicated by the figure. 
With GUT conditions at $5\%$ the large $\tan\beta$ solution for bottom--tau
Yukawa unification disappears because it does not predict an acceptable value 
for $V_{cb}$. The large $\tan\beta$ solution reappears if we relax the GUT 
conditions, and it is fully present imposing them at $40\%$.

\begin{figure}[htb]
\vspace{0.cm}
\centerline{ \hskip 0.cm \epsfxsize 3.0 truein \epsfbox {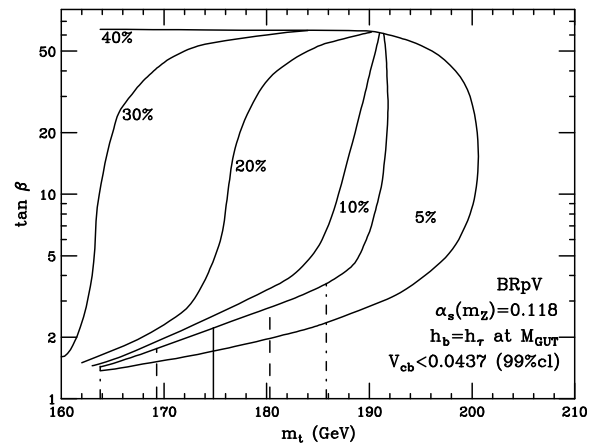}} 
\vskip -0.7cm
\caption{Regions with acceptable values of $V_{cb}$ in the 
$\tan\beta-m_{top}$ plane for BRpV.}
\vskip -0.7cm
\label{tl99Br}
\end{figure}
In Fig.~\ref{tl99Br} we present the same analysis but for BRpV. First of all, 
the allowed region corresponding to GUT conditions at $5\%$ is slightly larger
in BRpV. Indeed if we look only the region where $m_t$ is within $2\sigma$ 
of its experimental determination, the BRpV region is twice as large as in 
the MSSM \cite{textura}. More important differences between the two models 
appear when the GUT relations are relaxed: in BRpV it reappear the whole 
plane $m_t-\tan\beta$ as an allowed region when we relax the GUT conditions 
to $40\%$. 

The fact that $V_{cb}$ predictions tend to cut solutions with high values 
of $\tan\beta$ (observed in the MSSM as well as BRpV) can be understood
considering the RGE for the ratio between $|V_{cb}|$ and 
$R_{c/t}\equiv h_c/h_t$ \cite{BBO}. Imposing the relation in 
eq.~(\ref{VcbGUT}) at the GUT scale, we obtain at the weak scale
\begin{equation}
{{R_{c/t}}\over{|V_{cb}|^2}}(m_{w})\approx exp\left[{1\over{16\pi^2}}
\Big(h_t^2-h_b^2\Big)\ln{{M_{GUT}}\over{m_{weak}}}\right]
\label{RctVcbWeak}
\end{equation}
Since the left hand side of eq.~(\ref{RctVcbWeak}) is greater than one 
(approximately equal to 1.5), it is clear that the GUT condition 
$R_{c/t}=|V_{cb}|^2$ prefers the region of parameter space where the top 
Yukawa coupling is large while the bottom Yukawa coupling is small. This 
is obtained at small values of $\tan\beta$.

BRpV has many other important phenomenological consequences. The most crucial 
one for collider physics is the decay of the lightest supersymmetric particle
(LSP). This modifies all serach strategies for supersymmetric partners.
Here we mention also constraints from the decay mode $b\rightarrow s\gamma$.
It was shown that the constraints on the charged Higgs mass from the CLEO 
measurement for $B(b\rightarrow s\gamma)$ \cite{CLEObsg} in the MSSM 
\cite{bsgMSSM} are relaxed in BRpV \cite{bsgBRpV}.

In summary, BRpV, which provides an explanation for the solar and atmospheric 
neutrino problems, achieves $b-\tau$ unification at any value of $\tan\beta$,
predicts a value for $\alpha_s$ closer to the experimental value, and 
predicts the value of $V_{cb}$ in a wider region of parameter space, making 
it a serious alternative to the MSSM.

\section*{Acknowledgments}

I am thankful to  my collaborators A. Akeroyd, M.A. Garcia-Jare\~no, 
M. Hirsch, W. Porod, D.A. Restrepo, J.C. Rom\~ao, E. Torrente-Lujan, 
J.W.F. Valle, and specially J. Ferrandis, without whom this work would not 
have been possible.


\begin{thebibliography}{9}

\bibitem{brpvnew}
O.C.W. Kong, {\sl Mod. Phys. Lett.} {\bf A14}, 903 (1999); B. Mukhopadhyaya, 
S. Roy, and F. Vissani, {\sl Phys. Lett. B} {\bf 443}, 191 (1998); 
V. Bednyakov, A. Faessler, and S. Kovalenko, {\sl Phys. Lett. B} {\bf 442},
203 (1998); E.J. Chun, S.K. Kang, C.W. Kim, and U.W. Lee, {\sl Nucl. Phys. 
B} {\bf 544}, 89 (1999); A. Faessler, S. Kovalenko, and F. Simkovic, 
{\sl Phys. Rev. D} {\bf 58}, 055004 (1998); A.S. Joshipura, S.K. Vempati, 
hep-ph/9808232; C.-H. Chang and T.-F. Feng, hep-ph/9901260; D.E. Kaplan 
and A.E. Nelson, hep-ph/9901254; T.-F. Feng, hep-ph/9806505.

\bibitem{brpvus}
M.A. D\'\i az, D.A. Restrepo, J.W.F. Valle, hep-ph/9908286;
J. Ferrandis, hep-ph/9810371;
A. Akeroyd, M.A. D\'\i az, and J.W.F. Valle, {\sl Phys. Lett. B}
{\bf 441}, 224 (1998);
A. Akeroyd \etal, {\sl Nucl. Phys. B} {\bf 529}, 3 (1998).

\bibitem{SponRpB}
A. Masiero, J.W.F. Valle, {\sl Phys. Lett. B} {\bf 251}, 273, (1990);
J.C. Rom\~ao, C.A. Santos, J.W.F. Valle, {\sl Phys. Lett. B} {\bf
288}, 311 (1992); J.C.~Romao, A.~Ioannisian and J.W.~Valle, Phys.\
Rev.\ {\bf D55}, 427 (1997); J.C. Rom\~ao, J.W.F. Valle, {\sl
Nucl. Phys. B} {\bf 381}, 87 (1992); M.~Shiraishi, I.~Umemura and
K.~Yamamoto, Phys.\ Lett.\ {\bf B313}, 89 (1993); D.~Comelli,
A.~Masiero, M.~Pietroni and A.~Riotto, Phys.\ Lett.\ {\bf B324}, 397
(1994)

\bibitem{rpre}
G. G. Ross, J.W.F. Valle, {\sl Phys. Lett.} {\bf 151B} 375 (1985);
J. Ellis, G. Gelmini, C. Jarlskog, G.G. Ross, J.W.F.  Valle, {\sl
Phys. Lett.} 150B:142,1985; C.S. Aulakh, R.N. Mohapatra, {\sl
Phys. Lett. } {\bf B119}, 136 (1982); L.J. Hall and M. Suzuki, {\sl
Nucl. Phys. B} {\bf 231}, 419 (1984); I.-H. Lee, {\sl Phys. Lett. B}
{\bf 138}, 121 (1984); {\sl Nucl. Phys. B} {\bf 246},120 (1984);
A. Santamaria, J.W.F. Valle, {\sl Phys. Rev. Lett.} 60 (1988) 397 and
Phys. Lett. 195B:423,1987.

\bibitem{SKK}
SuperKamiokande Coll., Y. Fukuda {\it et al.}, {\sl Phys. Rev. Lett.} 
{\bf 81}, 1562 (1998) and {\sl Phys. Rev. Lett.} {\bf 81}, 1158 (1998).

\bibitem{expothers}
IMB Coll., R. Becker-Szendy \etal, {\sl Proc. Suppl. Nucl. Phys. B} {\bf 38}, 
331 (1995); Soudan Coll., W.W.M. Allison \etal, {\sl Phys. Lett. B} {\bf 449}, 
137 (1999); MACRO Coll., M. Ambrosio \etal, {\sl Phys. Lett. B} {\bf 434}, 
451 (1998).

\bibitem{solar}
Homestake Coll., B.T. Cleveland \etal, {\sl Aphys. J.} {\bf 496}, 505
(1998); GALLEX Coll., W. Hampel \etal, {\sl Phys. Lett. B} {\bf 447}, 127
(1999); SAGE Coll., J.N. Abdurashitov \etal, astroph/9907113; SuperK.
Coll., Y. Fukuda \etal, {\sl Phys. Rev. Lett.} {\bf 81}, 1158 (1998).

\bibitem{MSW}
S.P. Mikheyev and A.Yu. Smirnov, {\sl Yad. Fiz.} {\bf 42}, 1441 (1985)
[{\sl Sov. J. Nucl. Phys.} {\bf 42}, 913 (1985)] and {\sl Il Nuovo Cim.} 
{\bf C9}, 17 (1986); L. Wolfenstein, {\sl Phys. Rev. D} {\bf 17}, 2369
(1978).

\bibitem{RDHPV}
J.C. Rom\~ao \etal, hep-ph/9907499.

\bibitem{HirschV}
M. Hirsch and J.W.F. Valle, hep-ph/9812463.

\bibitem{Hempfling}
R. Hempfling, {\sl Nucl. Phys. B} {\bf 478}, 3 (1996). 

\bibitem{DRV}
M.A. D\'\i az, J. Rom\~ao, and J.W.F. Valle, {\sl Nucl. Phys. B} 
{\bf 524}, 23 (1998). 

\bibitem{DHaberii}
M.A. D\'\i az and H.E. Haber, {\sl Phys. Rev. D} {\bf 46}, 3086 (1992).

\bibitem{LEPmh} 
ALEPH Collaboration, hep-ex/9908016. 

\bibitem{textura}
M.A. D\'\i az, J. Ferrandis, and J.W.F. Valle, hep-ph/9909212.

\bibitem{topex} 
CDF Coll., F. Abe \etal, {\sl Phys. Rev. Lett.} {\bf 74}, 2626 (1995);
D0 Coll., S. Abachi \etal, {\sl Phys. Rev. Lett.} {\bf 74}, 2632 (1995).

\bibitem{yukunif}
M.A. D\'\i az, J. Ferrandis, J.C. Rom\~ao, J.W.F. Valle, {\sl Phys. 
Lett. B} {\bf 453}, 263 (1999). 

\bibitem{DFRV}
M.A. D\'\i az, J. Ferrandis, J.C. Rom\~ao, J.W.F. Valle, hep-ph/9906343.

\bibitem{wa}
C. Caso \etal, {\sl Eur. Phys. J.} {\bf C3}, 1 (1998).

\bibitem{alfamssm}
P. Langacker and N. Polonsky, {\sl Phys. Rev. D} {\bf 47}, 4028 (1993);
M. Carena, S. Pokorski and C.E.M. Wagner, {\sl Nucl. Phys. B} {\bf 406},
59 (1993);
P. Langacker and N. Polonsky, {\sl Phys. Rev. D} {\bf 52}, 3081 (1995).

\bibitem{expvcb}
Rev. Part. Phys., {\sl Eur. Phys. J.} {\bf C3}, 1 (1998). 

\bibitem{BBO}
V. Barger, M.S. Berger, and P. Ohmann, {\sl Phys. Rev. D} {\bf 47},
1093 (1993).

\bibitem{CLEObsg} 
CLEO Collaboration (M.S. Alam \etal), {\sl Phys. Rev. Lett.} {\bf 74},
2885 (1995).

\bibitem{bsgMSSM} 
J.L. Hewett, {\sl Phys. Rev. Lett.} {\bf 70}, 1045 (1993);
M.A. D\'\i az, {\sl Phys. Lett. B} {\bf 304}, 278 (1993) and 
{\sl Phys. Lett. B} {\bf 322}, 207 (1994);
K. Chetyrkin, M. Misiak, and M. M\"unz, {\sl Phys. Lett. B} {\bf 400}, 
206 (1997); M. Misiak, S. Pokorski, J. Rosiek, hep-ph/9703442 .

\bibitem{bsgBRpV}
M.A. D\'\i az, E. Torrente-Lujan, and J.W.F. Valle, {\sl Nucl. Phys. B} 
{\bf 551}, 78 (1999); 
E. Torrente-Lujan, hep-ph/9907220.

\end{thebibliography}
\end{document}